\documentclass[iop,revtex4-1]{emulateapj}
\usepackage{natbib}
\usepackage{hyperref}
\usepackage{newtxtext,newtxmath}
\usepackage{scalerel}
\usepackage{amsmath}	
\usepackage{amssymb}	
\usepackage{threeparttable}
\usepackage[dvipsnames]{xcolor}
\newcommand{\angstrom}{\textup{\AA}}
\usepackage{lineno}

\graphicspath{{./}{figures/}}
\newcommand{\mytilde}{\raise.19ex\hbox{$\scriptstyle\sim$}}

\begin{document}

\title{Discovery of a double radio relic in ZwCl1447.2+2619: A rare testbed for shock acceleration models with a peculiar surface brightness ratio}

\shorttitle{Double radio relic in ZwCL1447}
\shortauthors{Lee et al.}

\author{Wonki Lee\altaffilmark{1}}
\author{M. James Jee\altaffilmark{1,2}}
\author{Kyle Finner\altaffilmark{1,3}} 
\author{Kim HyeongHan\altaffilmark{1}}
\author{Ruta Kale\altaffilmark{4}}
\author{Hyein Yoon\altaffilmark{1,5,6}}
\author{William Forman\altaffilmark{7}}
\author{Ralph Kraft\altaffilmark{7}}
\author{Christine Jones\altaffilmark{7}}
\author{Aeree Chung\altaffilmark{1}}

\altaffiltext{1}{Department of Astronomy, Yonsei University, 50 Yonsei-ro, Seoul 03722, Korea; wonki.lee@yonsei.ac.kr, mkjee@yonsei.ac.kr}
\altaffiltext{2}{Department of Physics, University of California, Davis, One Shields Avenue, Davis, CA 95616, USA}
\altaffiltext{3}{Infrared Processing and Analysis Center, California Institute of Technology, Pasadena, CA 91125, USA}
\altaffiltext{4}{National Centre for Radio Astrophysics–Tata Institute of Fundamental Research, Ganeshkhind, Pune, Maharashtra, INDIA}
\altaffiltext{5}{Sydney Institute for Astronomy, School of Physics, A28, The University of Sydney, NSW 2006, Australia}
\altaffiltext{6}{ARC Centre of Excellence for All Sky Astrophysics in 3 Dimensions (ASTRO 3D)}
\altaffiltext{7}{Smithsonian Astrophysical Observatory, Harvard-Smithsonian Center for Astrophysics, 60 Garden St., Cambridge, MA 02138, USA}

\begin{abstract}
    We report a discovery of a double radio relic in the cluster merger ZwCl1447.2+2619 ($z=0.376$) with uGMRT observations at $420\rm~MHz$ and $700\rm~MHz$. The linear sizes of
    the northern and southern relics are $\mytilde0.3$~Mpc and $\mytilde1.2$~Mpc, respectively, which is consistent with
    the theoretical expectation that a larger relic is produced in the less massive subcluster side. 
    However, ZwCl1447.2+2619 is unlike other known double radio relic systems, where the larger relics are much more luminous by several factors. In this merger the higher surface brightness of the smaller northern relic makes its total radio luminosity comparable to that of the much larger southern relic.
    The surface brightness ratio $\mytilde0.1$ between the two radio relics differs significantly
    from the relation observed in other double radio relic systems.
    From our radio spectral analysis, we find that both relics signify 
    similar weak shocks
    with Mach numbers of $2.9\pm0.8$ and $2.0\pm0.7$ for the northern and southern relics, respectively.
    Moreover, the northern relic is connected to a discrete radio source with an optical counterpart, which indicates the possible presence of cosmic ray injection and
    re-acceleration.
    Therefore, we propose that this atypical surface brightness ratio can be explained with the particle acceleration efficiency precipitously dropping in the weak shock regime and/or with re-acceleration of fossil cosmic rays. Our multi-wavelength analysis and numerical simulation suggest that ZwCl1447.2+2619 is a post-merger, which has experienced a near head-on collision $\mytilde0.7\rm~Gyr$ ago.
\end{abstract}


\section{Introduction}


Mpc-scale diffuse radio emissions 
from merging galaxy clusters are broadly classified into two main categories: 
radio halos and radio relics. Radio halos are diffuse  structures 
roughly following the 
distribution of the intracluster medium \citep[ICM, e.g.,][]{Brunetti2014}. On the other hand, radio relics are characterized 
by 
arc-like morphologies 
in the cluster outskirts. 
As their name implies, radio relics 
have no 
distinct counterparts in other wavelength.
Also, their low spectral ages rule out the possibility that the relics originate from point source diffusion.
Thus, radio relics have 
been interpreted as indicating the 
locations of cosmic-ray formation (i.e., particle acceleration) 
\citep[e.g.,][]{1998A&A...332..395E,2010Sci...330..347V,VanWeeren2019}. 

It is now well accepted that the origin of radio relics is related to 
a cluster merger shock. Shocks can accelerate electrons via diffusive shock acceleration (DSA, e.g., \citealt{1983RPPh...46..973D}, \citealt{1987PhR...154....1B}) 
which generate high energy cosmic-ray electrons (CRes) with the Lorentz factors $\gamma\gtrsim10^3$, 
which emit a power-law synchrotron radiation 
($S_{\nu}\propto\nu^{-\alpha}$, e.g., \citealp{2012A&ARv..20...54F}). This model 
can explain 
the observed power-law spectrum at $100\rm~MHz <\nu < 10\rm~GHz$ \citep[e.g.,][]{Rajpurohit2018,2020A&A...636A..30R}. 
Discontinuities in the X-ray features across radio relics
have been reported 
in a few systems, which 
further 
support 
the shock-relic connection 
\citep[e.g.,][]{2013MNRAS.433.1701O,2015A&A...582A..87A,Botteon2016}.

The unique capability 
of radio relics 
to trace the merger shocks has been used to 
infer 
the history of the host cluster merger. To begin with, the 
very existence of a radio relic 
constrains the system to be a post-merger, which in general
is difficult to confirm with other observations alone.
Since radio relics propagate along the collision axis and are elongated perpendicular to it, the geometry of relics serves as a strong indicator of the past collision axis, even when the collision had a large impact parameter \citep[e.g.,][]{Lee2020}. 
Moreover, as the shocks are not 
decelerated by gravity, unlike the gas, dark matter, and galaxies,
the  distance to the radio relics from the cluster center can be used to estimate the time since the collision of the subclusters \citep[e.g.,][]{2019MNRAS.482...20Z,Kim2019,Hyeonghan2020}. 

Radio relics 
by themselves are also used as an astrophysical laboratory to
study plasma acceleration. 
Studies have 
demonstrated that the standard DSA alone is too inefficient in the weak shock regime ($\mathcal{M}\lesssim4$) to explain the observed luminosity of radio relics \citep[e.g.,][]{2020A&A...634A..64B}. 
Some studies 
have suggested a pre-acceleration mechanism, which enhances the DSA efficiency by pre-accelerating electrons
at 
$\mathcal{M}\gtrsim2.3$ with kinetic-scale instabilities formed in the upstream region \citep[e.g.,][]{2019ApJ...876...79K,2021ApJ...915...18H}. %
Other studies propose the presence of fossil CRes \citep[e.g.][]{2013MNRAS.435.1061P}. Fossil CRes are old, non-thermal populations generated by a past acceleration process. 
These supra-thermal ($\gamma\sim100$)  fossil CRes 
can be accelerated 
with much greater efficiency by weak shocks, compared to thermal electrons \citep[e.g.,][]{2011ApJ...734...18K}. 
This re-acceleration scenario has been 
supported 
by some observations, which 
show hints of re-acceleration from discrete sources \citep[e.g.,][]{Bonafede2014,Shimwell2015,Stuardi2019,2021MNRAS.505.4762J}. 
In particular, Abell 3411-3412 may present an 
on-going process of CRes injection from a cluster AGN to the radio relic \citep{VanWeeren2017}.

Clusters with double radio relics provide powerful tests of the different plasma acceleration models. 
In theory, every merger is expected to
create two merger shocks traveling in opposite directions.
However, to produce two observable radio relics, certain physical requirements should be met. Thus double radio relics are rare; a few tens have been reported to date (e.g., \citealp{Bonafede2009,2010Sci...330..347V,2011A&A...533A..35V,2012MNRAS.425L..36V,Kale2012,2015MNRAS.453.3483D,Stuardi2019}).
Since the two radio relics originate from the same collision, they share some initial conditions, such as the collision velocity. On the other hand, the properties of one relic can be significantly different from the other if affected by local environments such as CRes injection \citep[e.g.,][]{2021MNRAS.505.4762J}.
Therefore, the difference in their properties beyond the model prediction can be used to test different physical models in the relic formation. 

In this study, we report 
the discovery of double radio relic
in the 
galaxy cluster ZwCl1447.2+2619 (hereafter ZwCl1447) at $z=0.376$ \citep{2019ApJ...882...69G}. 
Since ZwCl1447 was identified by \cite{Butcher1984}
as the cluster with the highest fraction of blue galaxies of any cluster known at the time, ZwCl1447
has been a subject of a number of studies 
\citep[e.g.,][]{1999ApJ...524...22W,Giovannini2009,Govoni2012}.
ZwCl1447 was classified as a merging galaxy cluster by detection of diffuse radio emission and X-ray substructures from VLA and ROSAT observations, respectively \citep{Giovannini2009,Govoni2012}.

However, the limited resolution (FWHM$\sim0.5\arcmin$) in both X-ray and radio observations from the previous studies prevented accurate classification of the diffuse radio emissions and thus our understanding of the merger scenario in ZwCl1447. 
\cite{Giovannini2009} 
interpreted the northern radio emission as 
a collection of a halo and a relic whereas \cite{Govoni2012} regarded the entire northern radio emission as a radio halo and suggested that 
the southern diffuse radio emission is a relic. 
However, the claimed southern relic (only revealed by their subtraction of point sources) has an unusual dimension of $\mytilde0.5\rm~Mpc \times \mytilde1\rm~Mpc$ for a relic and is co-located with the bright radio point sources.

In this work, we performed low-frequency radio continuum observations in ZwCl1447 with the upgraded Giant Metrewave Radio Telescope (uGMRT). 
With the benefits of the wide frequency coverage at Band 3 ($250-500~\rm MHz$) and Band 4 ($550-850~\rm MHz$), 
we improved both resolution and sensitivity of the radio data by a factor of $\mytilde6$\footnote{Our band 4 image with robust parameter 0.5 and VLA-D image at $1.4\rm~GHz$ from \cite{Govoni2012} is compared.}.
Our interpretation is aided by the {\it Chandra} X-ray and Subaru weak-lensing (WL) observations, which are also first being presented.

The paper is organized 
as follows. 
The multi-wavelength observations and their reduction process are
described in \textsection\ref{sec:obs}. We present our main 
results in \textsection\ref{sec:result} 
and discussions in \textsection\ref{sec:discussion}
before we conclude in \textsection\ref{sec:summary}.  
In this paper, we use the $\Lambda$CDM cosmology with $H_0 =70 \rm\, km~s^{-1}~Mpc^{-1}$, $\Omega_{\rm m}=0.3$, and $\Omega_{\Lambda}=0.7$. In the adopted cosmology, $1\arcsec$ corresponds to $5.16\rm~kpc$ at the cluster redshift ($z=0.376$).

\begin{figure*}
    \centering
	\includegraphics[width=2\columnwidth]{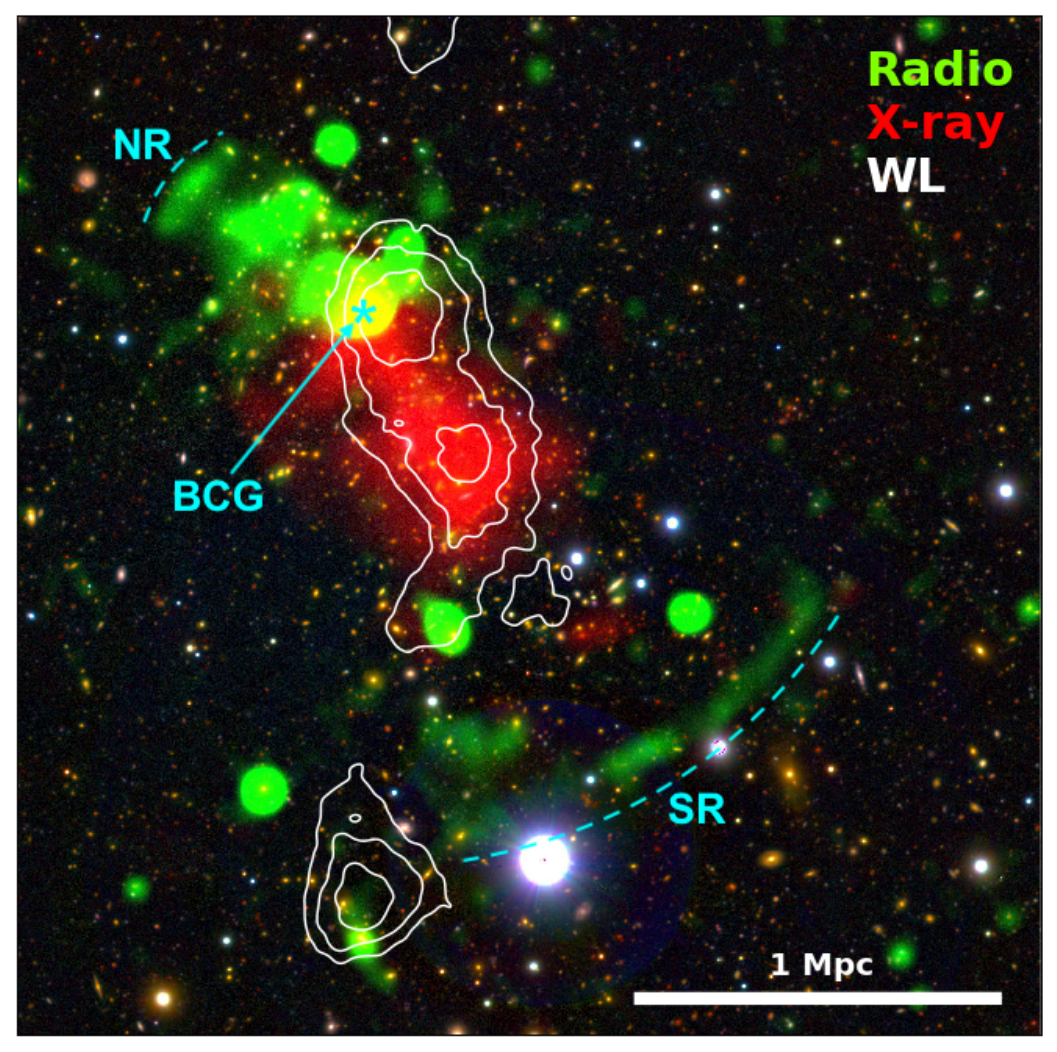}
    \caption{Multi-wavelength view of ZwCl1447 with uGMRT radio emission (green), \textit{Chandra} X-ray map (red), and the Subaru optical color composite image overlaid with the weak-lensing mass reconstruction contours (white). The radio emission is smoothed by $10\arcsec$ Gaussian kernel. The X-ray map is adaptively smoothed with \textit{csmooth} in {\tt ciao}$-$4.13. The effective smoothing kernel width of the WL mass reconstruction is $\mytilde40\arcsec$.  
    Our WL analysis detects two mass clumps coincident with the substructures identified from the X-ray observations. The brightest cluster galaxy (asterisk) is a strong radio point source and appears to belong to the northern subcluster. 
    Two elongated diffuse radio structures (NR and SR) are identified in the cluster outskirts with the orientation perpendicular to the elongation of the X-ray emission and the WL mass distribution. 
    }
    \label{fig:title}
\end{figure*}

\begin{figure*}
    \centering
	\includegraphics[width=2\columnwidth]{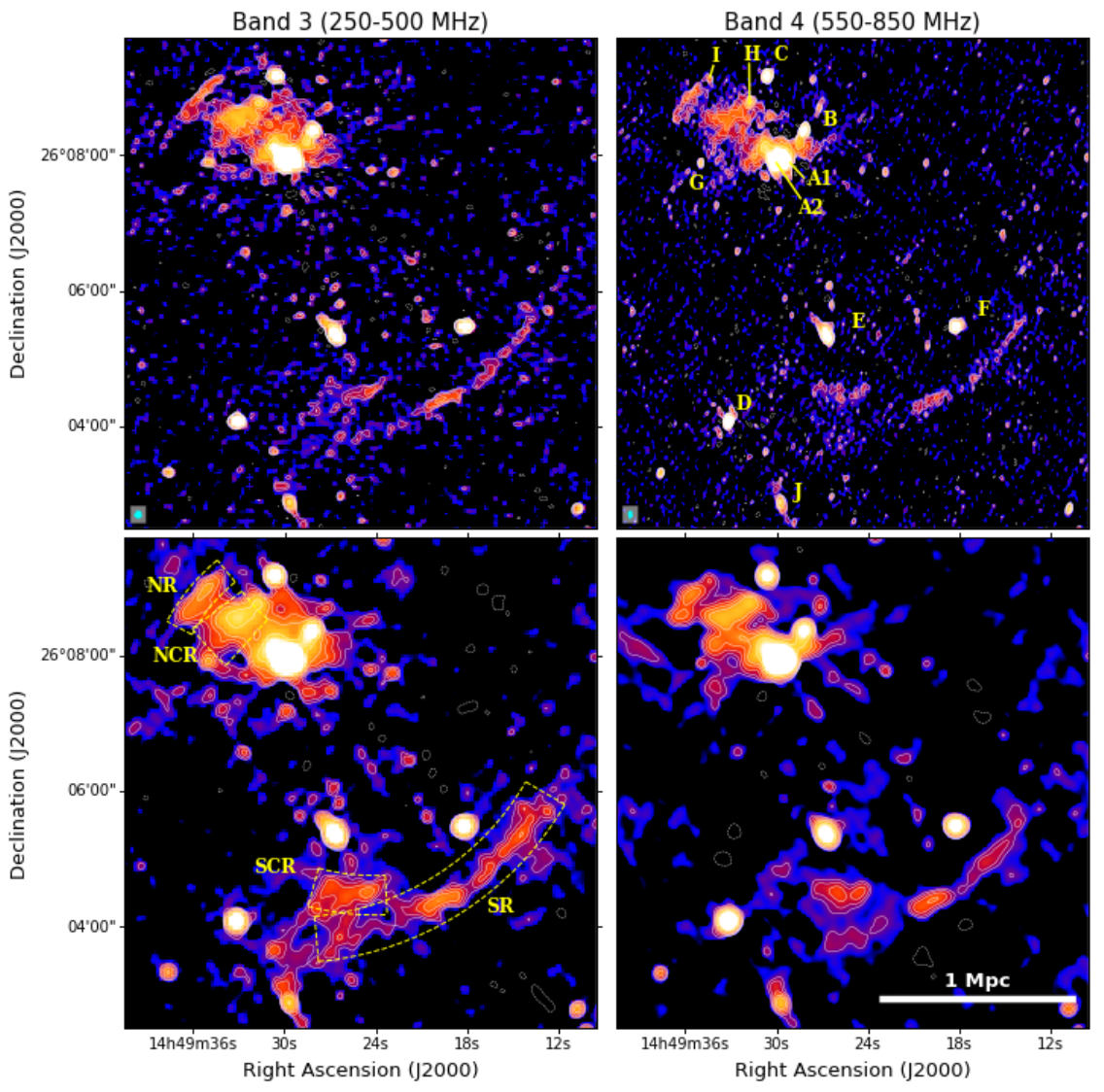}
    \caption{ZwCL1447 radio map at $420~\rm MHz$ (Band 3, $250-500~\rm MHz$, left) and $700~\rm MHz$ (Band 4, $550-850~\rm MHz$, right) with high- (top) and low-resolution (bottom) setups. The high- (low-)resolution images are cleaned with the robust parameter 0 (0.5) with Briggs weighting. 
    Synthesized beam of the high-resolution images is shown as the cyan ellipse in the bottom left corner. Low-resolution images are smoothed to have $10\arcsec$-size beam with {\tt imsmooth} to emphasize the diffuse radio emission.
    White solid contours mark the $3\sigma$ level and are spaced by a factor of $\sqrt{2}$. Dotted contours mark the $-3\sigma$ level. Noise in the high-resolution image is $\sigma_{\rm 420\, MHz,HR} = 28 \mu \rm Jy\,beam^{-1} (b=6\farcs7\times5\farcs6)$ and $\sigma_{\rm 700\, MHz,HR} = 18 \mu \rm Jy\,beam^{-1} (b=5\farcs7\times3\farcs7)$ and in the low-resolution image is $\sigma_{\rm 420\, MHz,LR} = 44 \mu \rm Jy\,beam^{-1}$ and $\sigma_{\rm 700\, MHz,LR} = 42 \mu \rm Jy\,beam^{-1}$. Annotations note the identified diffuse radio emissions (regions marked the yellow, dashed segments) and discrete radio sources. For NCR, we manually select the region that excludes the discrete source.}
    \label{fig:allradio}
\end{figure*}

\section{Observations}
\label{sec:obs}
\subsection{Radio: uGMRT}
\label{sec:ugmrt}
We observed ZwCL1447 using the uGMRT in Bands 3 ($250-500 \rm \,MHz$), 4 ($550-850 \rm \,MHz$), and 5 ($1000-1460 \rm \,MHz$) with on-source integrations of 2, 3, and 6 hours, respectively\footnote{Program code: $\rm 36\_002$ (Band 3, 4), $\rm 37\_124$ (Band 5).}.
The observations were carried out in 4096 frequency channels, with a sampling time of 5 seconds.
We chose the source 3C286 for both phase and amplitude calibration for all bands.
However, the Band 5 observation was severely affected by the 
correlator 
malfunction and 
thus we did not use Band 5 data for our 
scientific analysis.

We processed the uGMRT data with {\tt CASA v5.1.2-4}\footnote{http://casa.nrao.edu} using the  {\tt CAPTURE} pipeline \citep{2021ExA....51...95K}. We began our reduction by flagging the known bad channels and the first and last 10 seconds of each scan.
The remaining RFIs were flagged with the {\tt flagdata} task using the {\tt clip} and {\tt tfcrop} modes.
The first-round calibration was iteratively performed by 
computing the delay, gain, and bandpass solution. In the second round, we flagged the data by running {\tt flagdata} in the {\tt rflag} and {\tt tfcrop} modes, depending on the inclusion of arm antennas in the baseline.
In total, we flagged $\mytilde47\%$ and $\mytilde56\%$ of the data at Band 3 and Band 4, respectively. 
The calibrated data were then rebinned by grouping five channels, which gave channel widths of $\mytilde0.2\rm~MHz$ and $\mytilde0.5\rm~MHz$ for Bands 3 and 4, respectively. 

We ran the task {\tt tclean} with the options of the {\tt MS-MFS} \citep[multi-scale multi-frequency synthesis,][]{Rau2011} deconvolver, two Taylor terms ({\tt nterms}=2), and {\tt W-Projection} \citep{Cornwell2008} to accurately model the wide bandwidth and the non-coplanar field of view of uGMRT. To model the diffuse emission, we chose a set of scales that were $0,\,5,\,$ and $15\times$ the beam size for {\tt MS-MFS}.
The task {\tt tclean} automatically masks out regions during the cleaning process via the {\tt auto-multithresh} \citep{Kepley2020} algorithm. We visually inspected these masks before applying them.
We tested the ``briggs'' weighting scheme with different {\tt robust} parameters. We found that a {\tt robust} parameter of 0.5 best detects the diffuse radio emissions without losing much spatial resolution. 
These images are further smoothed when we need to match the beam size of radio images at different frequencies (\textsection \ref{sec:spectral}, \textsection \ref{sec:re-acc}). We confirmed that the flux of diffuse radio emissions is consistent in the radio image with a larger {\tt robust} parameter. We used a {\tt robust} parameter of 0 when resolving the compact radio emissions.
The resulting noise levels and beam sizes of the used cleaned images are summarized in Table 1.

We performed the primary beam gain correction with the task {\tt wbpbgmrt} \footnote{https://github.com/ruta-k/uGMRTprimarybeam}, which evaluates the primary beam corrections at multiple channels to properly account for the wide bandwidth of the uGMRT. 
We note that the effect of the correction is small as our target is concentrated on the image center ($\lesssim5'$).

We assume the flux calibration uncertainty is $10\%$ across all frequencies following \cite{2004ApJ...612..974C}. Then, the flux density uncertainty ($\Delta f$) 
is estimated from:
\begin{equation}
    \label{eq:error}
    \Delta f = \sqrt{(0.1f)^2 + N_{\rm beam}(\sigma_{\rm rms})^2},
\end{equation}
\noindent
where $f$ is the flux density, $N_{\rm beam}$ is the number of beams, and $\sigma_{\rm rms}$ is the background noise per beam. We 
compare the fluxes of the two 
discrete sources: FIRST J145007.4+254912 and FIRST J144803.1+262731 
with those from the publicly available catalogues \citep{Douglas1996,Lane2014}. We found consistent 
fractional differences of $15\%$ and $-20\%$ at $420 \rm \,MHz$ (Band 3) and $700 \rm \,MHz$ (Band 4), respectively, and applied the corrections to our data.

\subsection{X-ray: \textit{Chandra}}
\label{sec:chandra}
We 
investigate the X-ray properties of the cluster using the $40\rm~ksec$ \textit{Chandra} observations (ID: 20784, 22721, PI: R., Kraft), 
which were 
carried out with 
ACIS-I 
in VFAINT mode. 
Each program was processed with the \textit{chandra\_repro} script in {\tt ciao}$-$4.13. Then, we flagged and removed epochs with background flares by applying 2-sigma clipping with \textit{lc\_clean} in {\tt Sherpa}$-$4.13. The resulting exposure time is $30\rm~ksec$. We identified point sources with \textit{wavdetect} and masked them out after 
visual inspection.
The X-ray luminosity was estimated from the photons in the $[0.1,2.4]\rm~keV$ energy band and the temperature of each cluster was modeled by fitting the spectra obtained in the energy range  $[0.5,7.0]\rm~keV$ with the $phabs \times APEC$ model \citep[e.g.,][]{2006ApJ...640..710V}. 

\subsection{Optical: Subaru \& Keck/DEIMOS}
\label{sec:subaru}
We performed 
WL analysis using Subaru/SuprimeCam $g$, $r$, and $i$-band observations following the procedures of \cite{2020arXiv201002226F}. 
In brief, we modeled the PSF using principal component analysis on individual exposures and measured galaxy shapes through forward-modeling using PSF-convolved elliptical Gaussians.
Source galaxies were selected based on the color-color diagram by excluding the locus of the 91 spectroscopic members obtained from the Keck/DEIMOS and SDSS observations \citep{2019ApJS..240...39G}.
We achieved a source density of $\mytilde 30 ~\rm arcmin^{-2}$. 
Our mass reconstruction 
shows that ZwCl1447 
is comprised of at least three significant substructures (see Figure~\ref{fig:title}) that closely tracing the cluster galaxy distributions. We estimated the substructure masses by simultaneously fitting NFW models centered at individual BCGs. Readers are referred to Finner et al. (2021) for detailed descriptions on the individual steps. 

We generated a galaxy number density map using cluster member candidates. The 
candidates were first selected on the color-magnitude diagram with $19\leq m_r \leq 22$\footnote{The $r$-band magnitude of the BCG is $m_r=19$.} and 
$g-r \geq 1.2$.
We refined the selection based on the distribution of the spectroscopic members ($
\left|z-0.376\right| < 0.01$) on the color-color diagram. As a result, a total of 834 member candidates were selected and used to generate 
the cluster optical number density map. 

Our Keck/DEIMOS observations 
cover 
a wavelength range of 
$\mytilde4800 \angstrom$ to $\mytilde6800 \angstrom$ 
in the cluster rest frame. 
The wavelength coverage includes the $\rm H\alpha$, $\rm H\beta$, $\rm [OIII]$, and $\rm [NII]$ emission lines, which can 
be used to characterize the SF/AGN activities \citep[e.g.,][]{Baldwin1981}. 
We used the ratios of these emission lines to discuss the origin of the radio emission.



\begin{table*}
	\centering
	\caption{Properties of the cleaned uGMRT radio image.
	}
	\label{tab:image}
	\begin{tabular}{ccccccl} 
		\hline
		Band & Frequency\footnote{Frequency range that we used in the cleaning process.}  & On-source & Robust\footnote{Robust parameter used with briggs weighting.} & $\sigma_{\rm rms}$ & Resolution\footnote{Synthesized beam or smoothed beam size. Beam is smoothed to match the beam size between the images.}
		& Scientific usage\footnote{The scientific usage of each image and the name of the Figure that used the image.}\\
		& [MHz] & time (h)    & parameter & [$\mu \rm Jy\,b^{-1}$]   &  & \\
		\hline
		3 & 350-490  & 2.8 & 0 &28 & $6\farcs7 \times5\farcs7$ & High-resolution radio map (Fig. \ref{fig:allradio}) \\
		  &     &      & 0.5 & 31 & $8\farcs2 \times6\farcs6$ & 1D spectral index profile (Fig. \ref{fig:nr_steep}) \\
		  &     &      & 0.5 & 44 & $10\arcsec \times10\arcsec$ & Low-resolution radio map (Fig.  \ref{fig:allradio}), 2D spectral index map (Fig. \ref{fig:specmap}) \\
		4& 570-800  & 3.5 & 0 & 18 & $5\farcs7\times3\farcs7$ & High-resolution radio map (Fig. \ref{fig:allradio}) \\
		&     &     & 0.5 & 14 & $8\farcs2 \times6\farcs6$ &  1D spectral index profile (Fig. \ref{fig:nr_steep})\\
		 &     &    & 0.5 & 42 & $10\arcsec \times10\arcsec$ & Low-resolution radio map (Fig. \ref{fig:allradio}), 2D spectral index map (Fig. \ref{fig:specmap})\\
		\hline
	\end{tabular}\\
\end{table*}

\begin{table*}
	\centering
	\caption{Radio properties of diffuse radio emissions.}
	\label{tab:radioflux}
	\begin{tabular}{ccccccc} 
		\hline
		Name&  $S_{420~\rm MHz}$ & $S_{700~\rm MHz}$ & $LLS$\footnote{Largest linear size (LLS) of the diffuse radio emissions.} & $\alpha^{700\rm~MHz}_{420\rm~MHz}$ & $\mathcal{M}$ &  $P_{\rm1.4~GHz}$\footnote{Extrapolated luminosity at the rest frame frequency $1.4\rm~GHz$ assuming a power law.}\\
		 &  [mJy] & [mJy]   & [kpc] &  &  & [$10^{23} \rm\,W~Hz^{-1}$] \\
		\hline
		NR  & 4.38$\pm$0.48  & 2.30$\pm$0.26  & 300  &  1.27$\pm$0.31 &  2.9$\pm$0.8 & 5.1$\pm$2.2\\
		NCR & 11.67$\pm$1.19 & 5.30$\pm$0.55  & 340  &  1.55$\pm$0.29 & 2.2$\pm$0.7 & 10.6$\pm$4.2\\
		SR  & 13.70$\pm$1.44  & 5.81$\pm$0.64 & 1200 (1400\footnote{The LLS measured at $420\rm~MHz$.}) &  1.68$\pm$0.30 & 2.0$\pm$0.7 & 11.1$\pm$4.6\\
		SCR & 5.09$\pm$0.56  & 2.29$\pm$0.27  & 270  &  1.57$\pm$0.32 & 2.1$\pm$0.8  & 4.6$\pm$2.0\\
		\hline
	\end{tabular} \\
\end{table*}

\section{Results}
\label{sec:result}
Our multi-wavelength observations of ZwCl1447 are summarized in Figure~\ref{fig:title}.
The $Chandra$ X-ray 
observation reveals two X-ray substructures aligned in the NE-SW orientation (\textsection\ref{sec:xray}).
The WL mass 
reconstruction also presents two substructures 
coincident with the X-ray substructures (\textsection\ref{sec:wl_mass}), which strongly suggests that ZwCl1447 is a binary cluster merger.

The merger phase (pre-merger vs. post-merger) of the main binary structure is uncertain, based on the X-ray 
and 
WL data alone.
However, our discovery of the radio relics with uGMRT unambiguously 
confirms that the system is clearly in a post-merger state.
Two remarkable diffuse radio emissions are present in the cluster outskirts.
The vector connecting the centers of these two radio emissions is 
well-aligned with that defined by the aforementioned two substructures revealed by the WL and X-ray data, which 
is consistent with the expectation that the merger indeed happened in the NE-SW direction.
In addition, the elongations of both radio components are perpendicular to this hypothesized collision axis.
Based on their morphology, orientation, and location, we classify them as radio relics, which makes ZwCl1447 one of only a few tens of clusters that possess a double radio relic \citep{VanWeeren2019}. 

\subsection{Radio emissions in ZwCL1447}
\label{sec:radio}

Figure~\ref{fig:allradio} features the high-resolution (top, robust=0) and the low-resolution (bottom, robust=0.5) radio images at $420\rm~MHz$ (left) and $700\rm~MHz$ (right column). 
The two bottom-panel images were smoothed to have an equal beam size
of $10\arcsec\times10\arcsec$ to make the diffuse radio structures appear more prominent.
We describe the analysis of the annotated radio features in  \textsection\ref{sec:southern_relic}-4. Our spectral analysis is presented in \textsection\ref{sec:spectral}. 

\begin{figure}
    \centering
	\includegraphics[width=\columnwidth]{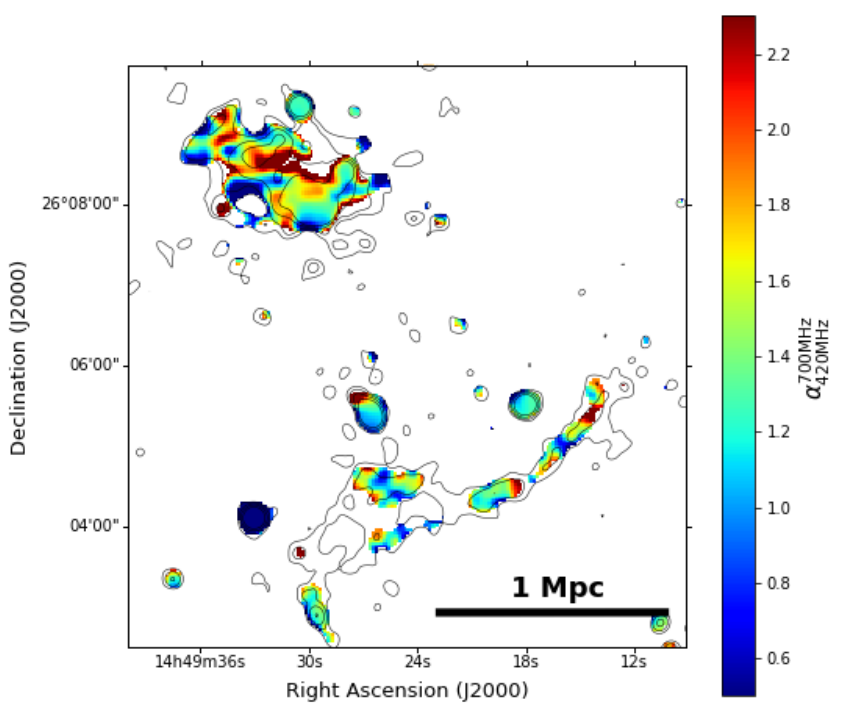}
	\includegraphics[width=0.98\columnwidth]{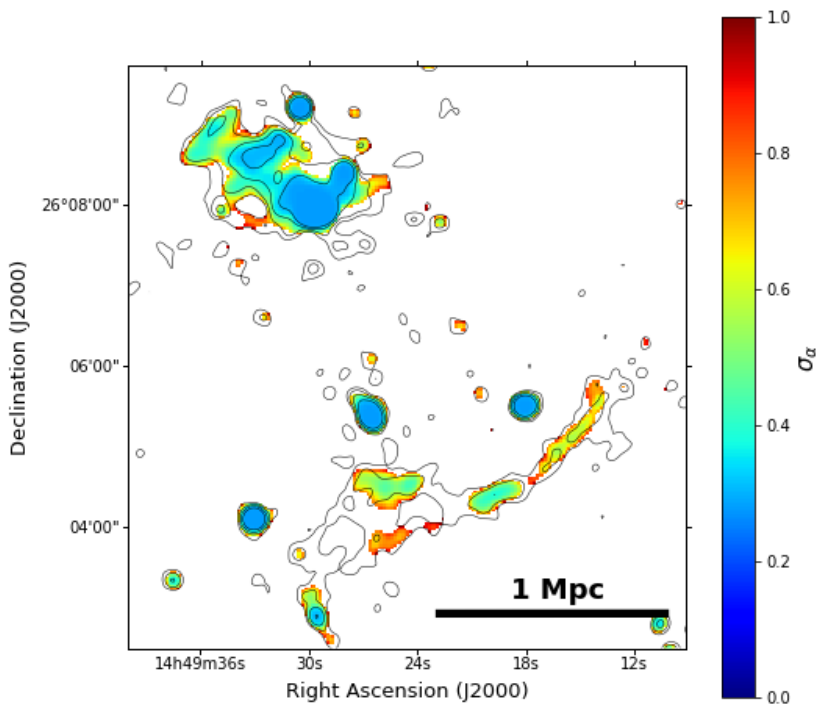}
    \caption{Spatial distribution of the spectral index (top) and its error (bottom) overlaid with the low-resolution $420\rm\,MHz$ radio contours. The spectral index is estimated by fitting a power law function for each pixel that exceeds $3\sigma$ on both $420\rm\,MHz$ and $700\rm\,MHz$ low-resolution radio maps. Spectral index and error map have a resolution of $10\arcsec\times10\arcsec$. 
    }
    \label{fig:specmap}
\end{figure}

\subsubsection{Southern radio relic}
\label{sec:southern_relic}
The southern radio relic (hereafter SR) is the largest radio structure in ZwCl1447. The 
extent of SR is $\mytilde 1.4\rm\, Mpc$ at $420\rm~\,MHz$ ($\mytilde 1.2\rm\, Mpc$ at $700\rm\,MHz$). 
SR is also remarkably thin. Using the contour level having
half the peak intensity in SR, we measure the average width to be $\mytilde 0.1\rm\, Mpc$.
SR is separated from the BCG by $\mytilde 1.3\rm\, Mpc$ and from the 
southern X-ray core by $\mytilde 0.8 \rm\,Mpc$. The latter is comparable to the virial radius of the southern cluster ($\mytilde 1\rm\, Mpc$, \textsection\ref{sec:wl_mass}). 
Such 
a thin, long morphology is rare and has been reported for only a few radio relics including the Sausage relic in CIZA J2242.8+5301 \citep[][]{2010Sci...330..347V}. We stress that this detailed structure was not accessible 
with the previous VLA data \citep[][]{Govoni2012}.

The surface brightness of SR varies along its 
extension. The $\mytilde 300\rm\, kpc$ central patch shows a factor of $\mytilde 2$ higher surface brightness than the wings. 
This resembles the feature seen in numerical simulations, where 
a spherical shock wave produces a center-bright radio relic \citep[e.g.,][]{2012MNRAS.425L..36V}. 
However, we caution that this interpretation should be viewed with caution, since in this region there also is a spectroscopic cluster member, which, if it provides a fossil CR cloud,
may explain the aforementioned brightness variation \citep[e.g.,][]{2021MNRAS.505.4762J}. 

\subsubsection{Northern radio relic}
\label{sec:northern_relic}
The northern radio relic (hereafter NR) 
is $\mytilde0.3\rm\,Mpc$ long and 
$\mytilde0.1\rm\,Mpc$ wide. The width is similar to that of SR.
Its distance ($\mytilde0.5\rm\,Mpc$) from the northern X-ray emission
rules out the classification of the feature as a radio halo.
The orientation is parallel to SR and perpendicular to the hypothesized merger axis. Also, as discussed in \textsection\ref{sec:spectral} and \textsection\ref{sec:re-acc}, 
there is an indication of width-wise (i.e., along the merger axis) spectral steepening.
Thus, we classify NR as a radio relic.

The western edge of NR appears to be connected to a discrete radio source (denoted as ``I" in the upper-right panel of Figure~\ref{fig:allradio}).
This radio source has an optical counterpart, whose photometric colors are consistent with those of the spectroscopic members (\textsection\ref{sec:point2}).
The connected feature hints at the interesting possibility that NR might have
originated from the AGN activity and shock-induced re-acceleration. We discuss this scenario in more detail based on the spectral profiles in \textsection\ref{sec:re-acc}.

\subsubsection{Other Diffuse Radio Emissions}
\label{sec:other_radio}
We identify two additional regions of diffuse radio emission.
We mark the feature near NR (SR) as NCR (SCR) in the lower-left panel of Figure~\ref{fig:allradio}. 
NCR was classified as a radio halo in previous studies \citep{Giacintucci2009,Govoni2012}. However, 
our high-resolution radio and X-ray data show that NCR is significantly offset ($\mytilde0.3\rm\, Mpc$) from the nearest X-ray emission (see Figure~\ref{fig:title}). Thus, we rule out that NCR is a radio halo.
SCR, which is less luminous than NCR, was not resolved in previous studies.
The morphological and geometrical features of NCR and SCR are similar. They are extended by $\mytilde0.3\rm\, Mpc$ in the direction nearly perpendicular to the hypothesized merger axis, as are the two radio relics. They are separated from their adjacent radio relics by similar distances ($\mytilde200\rm~kpc$). The thickness of SCR ($\mytilde0.15\rm\, Mpc$) is also similar to that of NCR ($\mytilde0.2\rm\, Mpc$); they are wider than SR and NR.
Therefore, 
it is plausible that 
they 
might have originated from similar mechanisms.

With our current data, their origin is unclear. One may suggest that NCR may have originated from a revived fossil plasma supplied by the neighboring radio galaxy (\textsection\ref{sec:point2}). However, NCR is also clearly seen at VLA $1.4\rm~GHz$ \citep[][]{Govoni2012}. Other known revived fossil plasma features usually have steep spectral indices and are hard to identify at high frequencies \citep[e.g.,][]{2020A&A...634A...4M}. A secondary merger shock is also possible \citep[e.g.,][]{2021A&A...651A..41C}.
Deeper data are needed for further analysis.

\subsubsection{Point-like radio sources}
\label{sec:point1}
Our uGMRT observations resolve many point-like radio sources (see the annotations 
in the upper-right panel of Figure~\ref{fig:allradio}). We measured their properties 
using the radio images whose PSFs are circularized and matched
(bottom panel in Figure~\ref{fig:allradio}).
The results are summarized in Table \ref{tab:pointsource}. We tested our flux measurement by comparing the extrapolated $1.4\rm~GHz$ flux with the result in \cite{Govoni2012} and verified that the fluxes of the discrete sources `B', `D', and `E' agree within their $1\sigma$ uncertainties. 

\subsection{Spectral analysis}
\label{sec:spectral}
Spectral analysis is a powerful tool to understand the origin of diffuse radio emissions. Under the assumption of DSA, a radio spectral slope can be translated into the shock strength. Furthermore, its spatial variation can 
inform us of the aging process. 
The shock strength ($\mathcal{M}$) satisfies the following equation: 
\begin{equation}
    \mathcal{M} = \sqrt{\frac{2\alpha_{\rm inj}+3}{2\alpha_{\rm inj}-1}},
    \label{eq:Mach}
\end{equation}
where $\alpha_{\rm inj}$ is the injection spectral index \citep[e.g.,][]{1983RPPh...46..973D}. 
In practice, it is difficult to obtain $\alpha_{\rm inj}$ directly and thus instead the integrated spectral index $\alpha_{\rm int}$ is measured.
Since the radio emission from  
newly accelerated electrons is mixed with the ones from (cooled) old populations, $\alpha_{\rm int}$ is larger than
$\alpha_{\rm inj}$.
For a planar shock, where the time since the last acceleration exceeds the cooling timescale, 
$\alpha_{\rm int}$ and 
$\alpha_{\rm inj}$ 
are related as follows  \citep[e.g.,][]{Kang2015a}:
\begin{equation}
    \alpha_{\rm int} \simeq \alpha_{\rm inj} + 0.5.
    \label{eq:alpha}
\end{equation}

\begin{figure*}
    \centering
	\includegraphics[width=1.95\columnwidth]{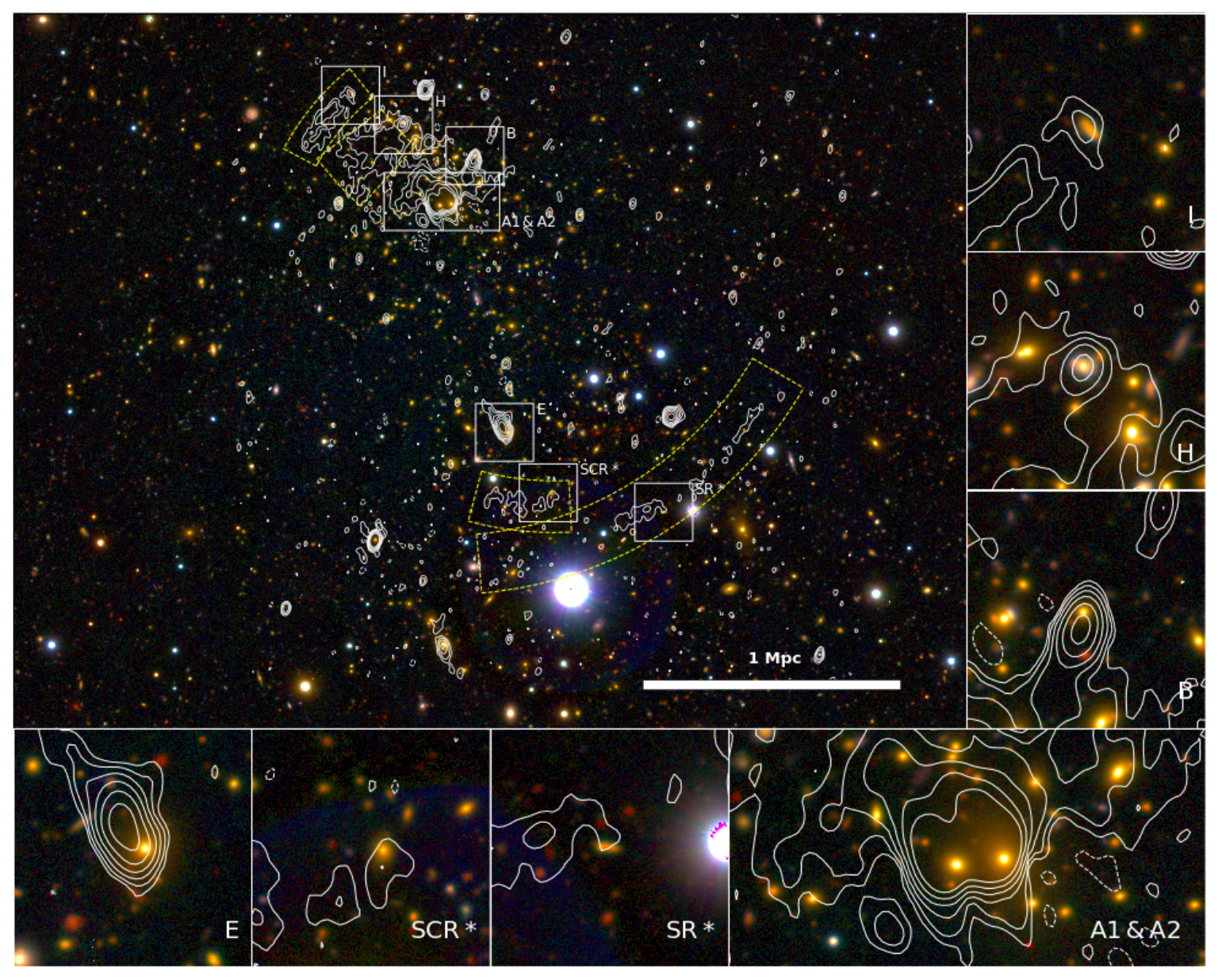}
    \caption{Subaru optical band ($g$, $r$, and $i$) color image of ZwCl1447 overlaid with the high resolution radio map at $700\rm~MHz$. The cropped images show the optical counterpart of the discrete radio source and the galaxies that are embedded in the diffuse radio emissions (`SCR*', `SR*'). Optical counterpart of `I' is classified as a photometric cluster member, whereas the remaining galaxies shown in the cropped images are spectroscopic members. Head-tail radio morphologies and galaxies embedded in the diffuse radio emission hint at the possibility that many of the local galaxies can inject fossil CRes into the diffuse radio structures. We signify AGN activity on `D' and SF activity on `H' and `SCR*' using the emission line ratio. We suspect that the SF activity of these cluster galaxy can be a source of fossil CRes injection on NCR and SCR.}
    \label{fig:galaxies}
\end{figure*}
We fit a power law 
to the integrated flux 
measured within the fan-shape regions in Figure~\ref{fig:allradio}. To account for
the difference in resolution between the two bands, 
we used the PSF-homogenized radio maps (bottom panel of Figure~\ref{fig:allradio}), which have the identical beam size of $10\arcsec$. 
The resulting flux measurements, integrated spectral indices, and Mach numbers are summarized in Table \ref{tab:radioflux}.

Our 
spectral analysis
indicates that the inferred shocks from the four diffuse emissions in ZwCl1447 are consistently weak, with the shock strength ($\mathcal{M}$) 
spanning a range from 2.0 to 2.9.
We verify that the spectral slopes for NR and NCR are consistent with what we obtain from the archival VLA data ($\alpha^{1.4\rm\,GHz}_{400\rm\,MHz}\sim1.3$ and $\sim1.5$ for NR and NCR, respectively)\footnote{The VLA data is processed using {\tt vlarun} in {\tt AIPS} and a power-law was fitted on the three frequencies data.}. Note that SR and SCR are not detected with the VLA data.


\begin{table}
	\centering
	\caption{Radio properties of discrete radio sources.}
	\label{tab:pointsource}
	\begin{tabular}{ccccc} 
		\hline
		Name&  RA & Dec &  $S_{420\rm~MHz}$ [mJy] & $S_{700\rm~MHz}$ [mJy] \\
		\hline
		A1\&A2 & 14\,49\,29.7 & 26\,07\,55.5 &  82.1$\pm$8.2  & 42.8$\pm$4.3  \\
		B  & 14\,49\,28.1 & 26\,08\,21.5 & 9.1$\pm$0.9 & 4.0$\pm$0.4   \\
		C  & 14\,49\,30.6 & 26\,09\,10.2 & 6.0$\pm$0.6 & 3.2$\pm$0.3    \\
		D  & 14\,49\,33.1 & 26\,04\,04.7 & 7.1$\pm$0.7 & 8.9$\pm$0.9   \\
		E  & 14\,49\,26.6 & 26\,05\,21.2 & 9.4$\pm$0.9 & 5.2$\pm$0.5    \\
		F  & 14\,49\,18.1 & 26\,05\,29.0 & 6.0$\pm$0.6 & 3.4$\pm$0.3  \\
		G  & 14\,49\,34.9 & 26\,07\,53.3 & 1.2$\pm$0.1 & 0.5$\pm$0.1  \\
		H  & 14\,49\,31.7 & 26\,08\,47.0 & 4.8$\pm$0.5 & 2.1$\pm$0.2  \\
		I  & 14\,49\,34.3 & 26\,09\,07.4 & 1.1$\pm$0.2 & 0.7$\pm$0.1  \\
		J  & 14\,49\,29.7 & 26\,02\,53.0 & 1.7$\pm$0.2 & 1.0$\pm$0.1 \\
        \hline
	\end{tabular}
\end{table}

\begin{figure*}
    \centering
	\includegraphics[width=2\columnwidth]{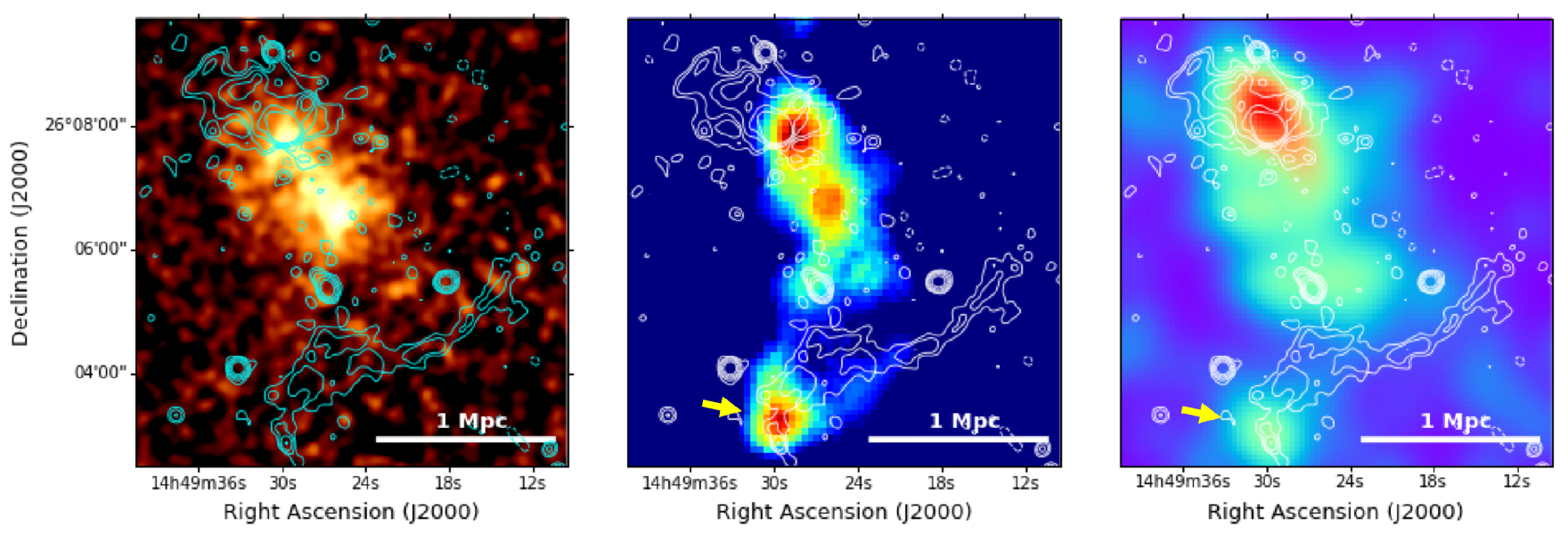}
    \caption{\textit{Chandra} X-ray map (left), weak-lensing mass map (middle), and photometric member galaxy number density map (right) overlaid with low-resolution $420\rm\,MHz$ radio contours. The X-ray image is the point source-subtracted, exposure corrected, and smooth with a $5\arcsec$ Gaussian. The effective smoothing kernel size of the WL mass reconstruction is $\mytilde40\arcsec$. All maps feature at least two substructures that are enclosed by the double radio relic and aligned in the NE-SW direction. The two substructures are most likely the components that collided and generated double radio relic. A third substructure can be identified in the southern outskirts in the WL mass map and the number density map (yellow arrow). The X-ray observation suffers from poor exposure time at the third substructure, due to the chip gap.}
    \label{fig:non_radio}
\end{figure*}

We present the spatial distribution of the spectral index and its error map in Figure~\ref{fig:specmap}. 
We limit our analysis to the regions where the significance is $>3\sigma$ at both frequencies. 
The spectral map of NR, if we exclude the region on the discrete source `I' that shows a large spectral index, 
hints at spectral steepening with the measurement at the northeastern edge (i.e., away from the cluster) being flatter  ($\alpha \sim 0.5$) than the value ($\alpha \gtrsim 2$) at the southwestern edge (i.e., toward the cluster).
This steepening direction is consistent with a merger shock direction, which we expect to propagate towards northeast along the merger axis and so accelerates particles on its leading edge at northeast.
Also, the map indicates the possible presence of longitudinal steepening away from the discrete source `I', which can be interpreted as the feature originating from the point source (e.g., CRes diffusion). 

In NCR, a flat spectrum is 
found near the discrete source `H' and the spectrum steepens toward the east. One may interpret this as CRes diffusion from the source `H' toward the east.
However, since we can also 
find a region with a flat spectrum near the eastern edge (furthest from the discrete source), it is difficult to explain the observed feature with this scenario alone. 

We do not see a clear spatial trend of spectral steepening in SR nor SCR where the spectral index is fluctuating lengthwise.
This complex spatial variation may be attributed 
to the possible presence of  
a turbulent medium \citep[e.g.,][]{Paola2021}
or large measurement errors.


\subsection{Radio-emitting cluster galaxies}
\label{sec:point2}
SF/AGN activity of cluster galaxies can supply CRes to ICM \citep[e.g.,][]{1996SSRv...75..279V,VanWeeren2017}. These CRes are visible in radio for only few tens of Myr by a radiative loss. 
Thus, the CRes injected in the past (i.e. fossil CRes) do not directly contribute to the cluster radio emissions. Nevertheless, they can provide a supra-thermal seed population and enhance the acceleration efficiency of merger shocks \citep[e.g.,][]{2011ApJ...734...18K,2013MNRAS.435.1061P}. Therefore, we can suspect a contribution of fossil CRes on those diffuse radio emissions near the cluster radio galaxies. 

In Figure~\ref{fig:galaxies}, we present the zoomed-in 
Subaru optical color images of the cluster galaxies that
co-locate with significant radio emissions.
As mentioned earlier, galaxy `I' is a photometric cluster member ($z_{\rm phot} = 0.4\pm0.1$, \citealp{2016MNRAS.460.1371B}) while the rest are spectroscopically confirmed members.
    
Many radio point sources show a head-tail morphology. Asymmetric tails of $100\rm~kpc$-scale 
are observed 
in `A1', `A2', `B', and `E'. 
And, except for `B', the tails are extended 
parallel to the hypothesized merger axis. 
Such alignments 
may be attributed to a strong bulk motion of the ICM along the merger axis 
\citep[e.g.,][]{1979ApJ...234..818J}. 
Under this merger configuration, it is plausible that
a significant amount of fossil CRes 
are fed 
into the cluster environment along the merger axis. 

Using the Keck/DEIMOS spectra, we 
searched for signs of SF/AGN activity. We derive the line ratios for spectra that posses strong emission lines by Gaussian fitting. The optical counterpart of `D' has a strong [OIII] emission line, signifying AGN activity. 
On the other hand, the targets `H' and `SCR$^*$' present strong $\rm {H}\alpha$ and $\rm {H}\beta$ emission lines 
over $\rm {[NII]}$ and $\rm {[OIII]}$, which  
satisfies the criteria 
for SF 
activities. 
Thus, we suspect that NCR and SCR, which 
harbor these galaxies, may originate from 
re-acceleration of fossil CRes injected by the SF activity of the cluster galaxies.


\subsection{X-ray emissions}
\label{sec:xray}
The left panel of Figure~\ref{fig:non_radio} shows the {\it Chandra} X-ray image of ZwCl1447 smoothed with a
$\sigma=5\arcsec$ Gaussian kernel. 
The highly elongated morphology of the X-ray map supports our NE-SW post-merger scenario.
The two X-ray peaks are well aligned with the two WL mass clumps (middle panel of Figure~\ref{fig:non_radio}).
Also, the 
northern X-ray 
peak is in good spatial agreement with the BCG, which
is approximately cospatial with the highest peak of the galaxy number density (right panel of Figure~\ref{fig:non_radio}).
On the other hand, the southern X-ray peak does not coincide with a distinct
galaxy overdensity.
Since our current spectroscopic catalog is highly incomplete, further investigation on this issue using spectroscopic members requires a deeper observation. 
  
We perform 
X-ray spectral analysis by fitting a plasma model as described in \textsection\ref{sec:chandra}. As a whole ($r<800\rm~kpc$), 
the 
temperature of the cluster is measured 
to be $4.5\pm0.8\rm~keV$. 
The X-ray 
luminosity is measured to be $L_X\sim3.66\times10^{44} \rm erg\,s^{-1}$ at $0.1\,-2.4\rm\,keV$,
which is consistent 
with the ROSAT-based result \citep{1999ApJ...524...22W, Govoni2012} and with the mass-derived luminosity (\textsection\ref{sec:wl_mass}, \citealt{2010A&A...511A..85P}). The spectrum for each substructure is derived from the $r=200\rm~kpc$ circular region centered on the peak.
We find that the northern and southern substructures 
have 
temperatures (X-ray luminosities) of $4.4\pm0.9\rm~keV$ ($\mytilde5\times10^{43} \rm erg\,s^{-1}$) and $3.3\pm0.5\rm~keV$ ($\mytilde9\times10^{43} \rm erg\,s^{-1}$), respectively. 
The current X-ray data does not allow us to detect any surface brightness discontinuity at the location of the radio relic.




\begin{figure*}
    \centering
	\includegraphics[width=1.9\columnwidth]{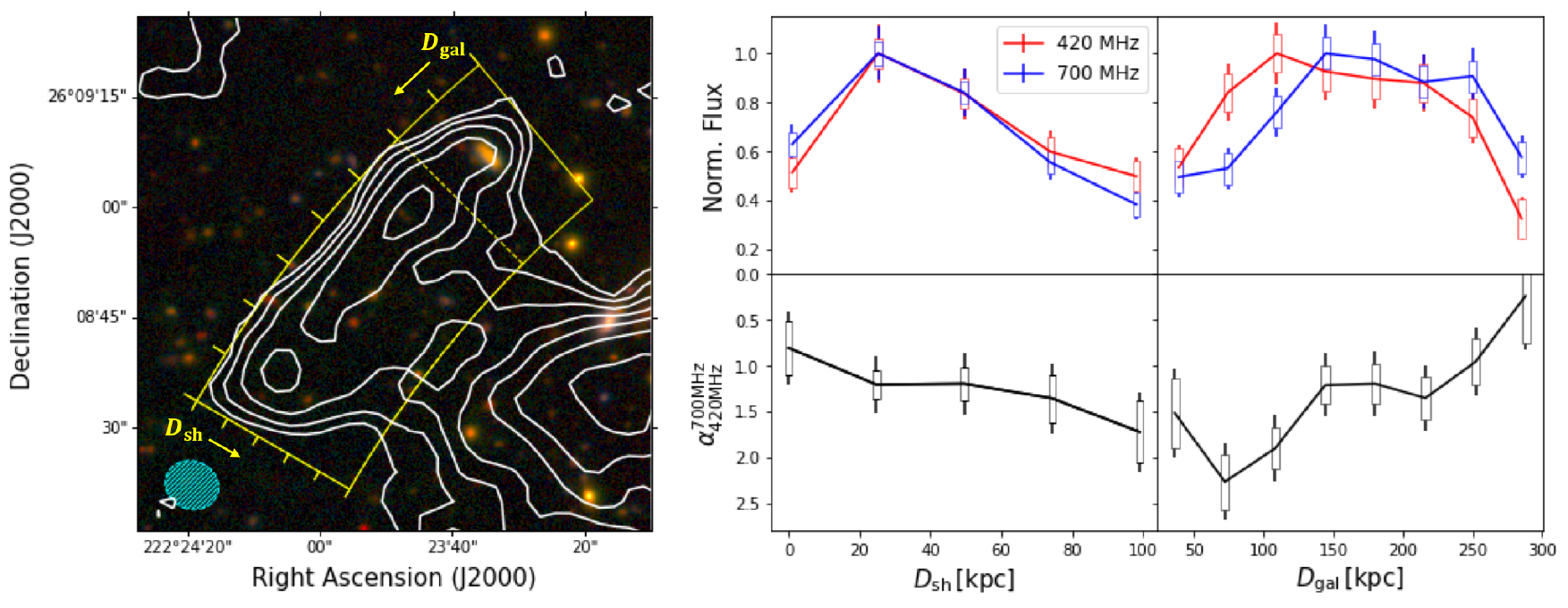}
    \caption{(Left) Cropped Subaru optical band image of NR overlaid with the $420\rm~MHz$ radio contours (robust=0.5). Flux (top) and integrated spectral slope profile (bottom row) of NR along radial (middle) and tangential direction (right column). Bins that are used to define the region of flux integration are marked with ticks along the fan-shaped region in the left figure. The reference point of the radial and the tangential profile is the shock surface (i.e. radial outermost bin) and the discrete source `I', respectively. `x'-axis of the radial and the tangential profile is the distance from the shock ($D_{\rm sh}$) and from the galaxy ($D_{\rm gal}$). We mark two errorbars: one that is based on the Equation \ref{eq:error} (line) and the other that does not count the flux calibration error (box). We use the latter in comparing the spectral index as the index will change throughout the entire profile with a zero-value correction by the flux calibration. The spectral slope in the radial spectral profile shows a hint of spectral steepening ($2.1\sigma$), which supports the shock acceleration scenario.}
    \label{fig:nr_steep}
\end{figure*}

\subsection{Optical band \& WL analysis}
\label{sec:wl_mass}

The WL mass and 
galaxy number density maps are highly correlated and both are extended in the same direction, along the hypothesized merger axis. However, as mentioned in \textsection\ref{sec:xray}, the galaxy number density
at the location of the southern WL and X-ray peaks is
not significantly concentrated. 

In addition to the main two (NW and SE) substructures, 
a third substructure is detected $\mytilde1.4\rm\,Mpc$ south of the BCG with both WL and galaxy number density maps (yellow arrow in Figure~\ref{fig:non_radio}). This substructure is not seen in the Chandra X-ray map. Considering the RASS X-ray overdensity visible at this location \citep{Govoni2012}, it is possible that the absence of the signal from the current {\it Chandra} data may be due to the coincidence of the CCD chip gap with the substructure. In this study, we assume that the third substructure did not contribute significantly to the formation of the current X-ray and radio features. With three-body encounter, it is difficult to explain the symmetric morphologies of the X-ray and radio emissions and the alignment of the two merger axes inferred from them individually.

A very bright star ($r\sim9$ mag) is present, $\mytilde 4 \arcmin$ south of the SE substructure. If left unmasked, many spurious sources whose shapes are tangentially aligned with respect to the star would produce a strong false WL peak at the location of the star. Thus, we applied a large circular mask to prevent it. Since the location of the star is distant from the main (NW and SE) substructures, the impact of this masking should be insignificant on our WL analysis. Alternatively, one can choose to address the issue by subtracting the stellar PSF profile and detecting sources near the star as is done for the analysis of A1240 in \cite{2021arXiv210906879C}. This involves careful identification and removal of spurious sources due to imperfect subtraction. We verified that the WL results when this method is applied remain unchanged.


The masses of these two substructures were derived by fitting two NFW profiles \citep{Navarro1996} simultaneously. We fix each profile center 
to the nearest BCG. 
The mass of the northern (southern) substructure is 
determined to be $2.7\pm0.8\times10^{14}M_{\rm sun}$ 
($1.0\pm0.5\times10^{14}M_{\rm sun}$). 
The result shows that ZwCl1447 is a major merger with a $\mytilde3:1$ mass ratio. 
Our single halo fitting with the center on the BCG gives 
a total mass of 
$\mytilde5\times10^{14}M_{\rm sun}$. 



\section{Discussion}
\label{sec:discussion}


\begin{figure*}
    \centering
	\includegraphics[width=2\columnwidth]{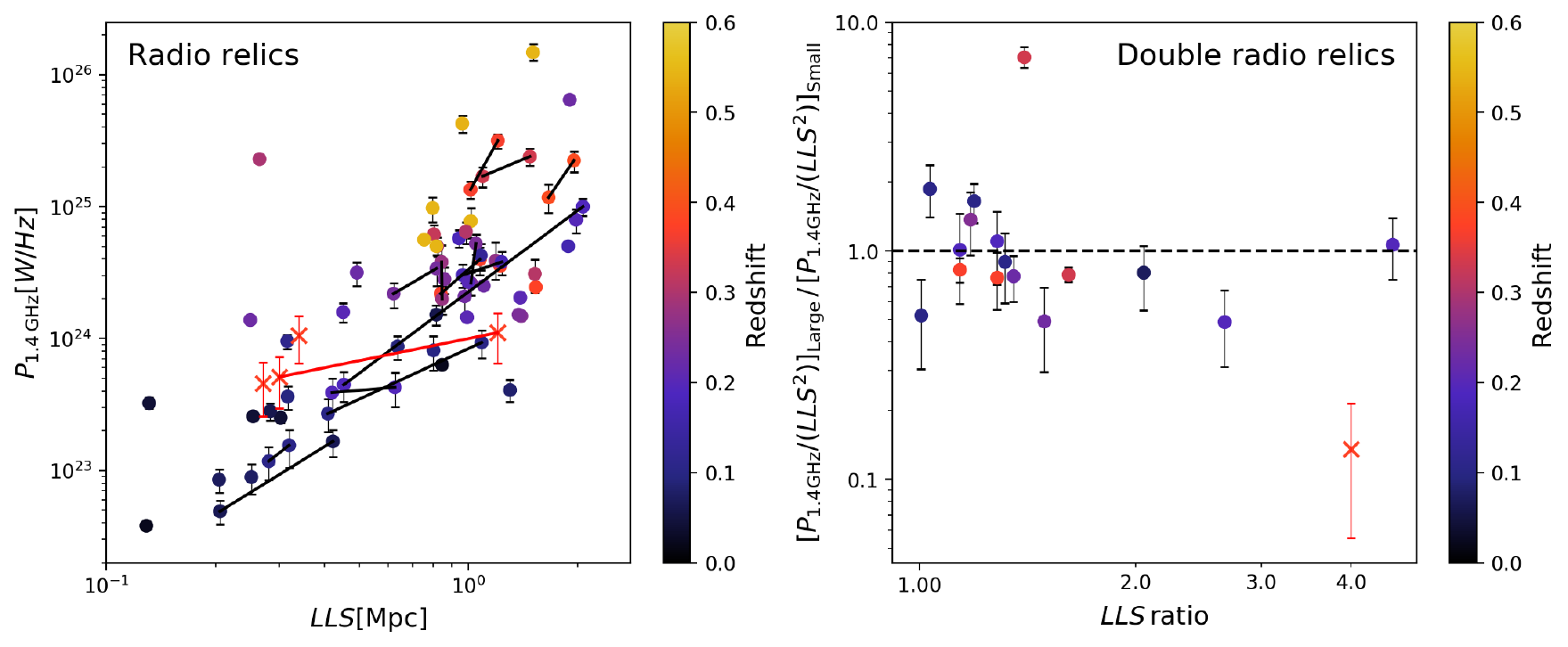}
    \caption{Size-luminosity relation and surface brightness ratio comparison of the ZwCl1447 radio relics with other known systems. Left panel: The largest linear size (LLS) and the radio luminosity at $1.4\rm~GHz$ ($P_{1.4\rm~GHz}$). We display NCR and SCR as well as the double radio relic in ZwCl1447 (red cross). Filled circles (colors represent the redshift of the host cluster) are other radio relics that we compiled from the literature \citep{VanWeeren2019,Kale2012,2003PhDT.........3J,2014MNRAS.444.3130D,2020MNRAS.496L..48L}.
    The solid lines connect the double relics within the same system.  
    The diffuse radio emissions in ZwCl1447 follow the size-luminosity relation seen in other systems. 
    Right panel: The LLS and surface brightness ratios at $1.4\rm~GHz$ of the double radio relics. The surface brightness is computed from the assumption that the the shock surface area is proportional to the square of LLS. ZwCl1447 (red cross) is a significant outlier showing a very low surface brightness ratio ($\mytilde0.1$) while most other double radio relic systems have a surface brightness ratio near unity.}
    \label{fig:Relation}
\end{figure*}

\subsection{Possible evidence of re-acceleration?}
\label{sec:re-acc}
In \textsection\ref{sec:northern_relic} and \textsection\ref{sec:spectral}, we presented the possibility of 
re-acceleration in NR based on the presence of the discrete source and the two-dimensional (2D) spectral map. To further 
investigate the possibility, we generate one-dimensional (1D) profiles of integrated spectral index 
in two directions: one along the extension from the adjacent discrete source to the opposite end (lengthwise) and 
the other along the hypothesized merger axis from the northern edge to the southern edge (widthwise). These 1D profiles 
provide a higher S/N diagnostic on the impact of the discrete radio source (lengthwise) and the shock-induced acceleration (widthwise). 
For this 1D profile measurement, unlike the case for the 2D spectral map generation, we used the high-resolution $8\farcs2\times6\farcs6$ images (see Table 1) produced with a robust parameter of 0.5 because we desire to minimize the flux mixing between two neighboring bins; we degraded the Band-4 image resolution slightly with a small difference kernel, derived with {\tt pypher}  \citep{2016A&A...596A..63B}, to match the Band-3 PSF.


Figure~\ref{fig:nr_steep} shows the 
resulting 1D profiles in NR. 
As 
indicated by the 2D spectral map (Figure~\ref{fig:specmap}), 
spectral steepening is present in the profile along the merger axis (left panel in
Figure~\ref{fig:nr_steep}). 
The spectral index is flattest 
($0.8\pm0.3$) at the northern edge and 
steepens toward the southern edge, reaching 
the highest value ($1.7\pm0.3$) at the 
southern edge. 
Thus, the 1D profile along the merger axis shows a hint of spectral steepening ($\mytilde2.1\sigma$), which supports the shock acceleration.

Along the lengthwise profile, we 
find a spectral flattening from the discrete source to the opposite end. 
This might indicate that the energy of the seed CRes of NR varies lengthwise. However, the observed pattern is opposite to the one 
expected in a re-accelerated jet,
which in general should show spectral steepening (thus aging) away from the AGN \citep[e.g.,][]{VanWeeren2017}.
Perhaps, this flattening 
can be explained with a tilted geometry of the jet. If 
the jet is injected with a non-zero angle with respect to the shock surface, the time since the last acceleration can vary lengthwise; 
in other words, the eastern edge can show a flat spectral index as observed if 
it is re-accelerated more recently than other regions \citep[e.g.,][]{Cuciti2018}. 
However, to further investigate this scenario, detection of spectral curvature 
at higher frequencies is required \citep[e.g.,][]{2016ApJ...823...13K}.

\subsection{Why is the southern relic so dim?}
\label{sec:luminosity}
An empirical relation between the size and luminosity of radio relics has been established by many previous studies \citep[e.g.,][]{2014MNRAS.444.3130D, VanWeeren2019}. Assuming a power-law spectrum of the radio sources, we extrapolate our measurements and derive the $k$-corrected luminosity at $1.4\rm~GHz$ using the following equation: 
\begin{equation}
    \label{eq:Lum}
    L_{1.4\rm~GHz} = \frac{4\pi D^2_{\rm L}(z)}{(1+z)^{1-\alpha}}\left( \frac{1.4\rm~GHz}{\nu} \right )^{-\alpha}S_\nu,
\end{equation}
\noindent
where $D_L(z)$ is the luminosity distance at the redshift $z$.
The derived values are presented in Table \ref{tab:radioflux}. The combined luminosity of NR and NCR are lower than the measurement of \cite{Govoni2012}, where the two sources were unresolved.
We attribute this discrepancy to the removal of the discrete radio sources which are resolved in the current study and excluded when we define the area for flux integration.

We compare the largest linear size and luminosity of the ZwCl1447 radio relics with those of other radio relics from the literature
in the left panel of Figure~\ref{fig:Relation}. 
Each radio relic in ZwCl1447 follows the size-luminosity relation of other known systems.
In the same figure, we link the double radio relic from the same system with a solid line whose slope then represents the relative property within the same system. We find that the ZwCl1447 slope is considerably flat when compared with those of other systems.
The contrast increases when we compare the surface brightness ratios (right panel of Figure~\ref{fig:Relation}); here, we used $\mytilde LLS^2$ as a proxy for the
surface area of each radio relic.
The majority of the double radio relic systems shows a surface brightness ratio near unity. This includes the Sausage radio relic system, which is the rightmost data point (LLS ratio$\sim 4.6$). 
However, the double radio relic system of ZwCl1447 is a significant outlier in this relation as the size ratio is large ($\mytilde4$) whereas the surface brightness of SR is an order of magnitude fainter ($\lesssim0.1$).

It is difficult to reconcile this peculiar surface brightness ratio 
with the subcluster mass ratio found in our WL analysis, which
shows that the northern cluster is $\mytilde3$ times more massive.
%
According to numerical simulations, 
the radio emissivity of the 
merger shock created by the less massive subcluster is higher and thus its radio relic is larger and brighter \citep[e.g.,][]{Ha2018,Lee2020}. 

The surface brightness 
of radio relics depends also on 
the cluster environment and the shock strength. 
Among these parameters, we assume that the upstream gas properties (i.e., density, temperature) and the magnetic field strength are similar between the two radio relics, which is plausible as the X-ray properties of the two substructures and the widths of the two radio relics are similar. 
According to \cite{Hoeft2007}, the surface brightness ratio depends on $\xi_{\rm e}$ (the fraction of the kinetic energy converted to the non-thermal component) and $\Psi(\mathcal{M})$ (the combined strength of the shock). 
$\Psi(\mathcal{M})$ is a sensitive function of $\mathcal{M}$, which rises sharply from a negligibly small value to unity in the $2<\mathcal{M}<4$ regime.
Moreover, recent numerical studies have suggested that only the shocks stronger than $\mathcal{M}_{\rm crit}\sim2.3$ can build kinetic-scale instabilities, which are the main drivers of electron pre-acceleration \citep[e.g.,][]{2021ApJ...915...18H}.
Thus, under this simplified assumption, these sensitivities might explain the peculiar surface brightness ratio between the two relics in ZwCl1447 because the observed $\mathcal{M}$ values ($2.9\pm0.8$ and $2.0\pm0.7$ for the northern and southern relics, respectively) lie in the aforementioned critical range ($2<\mathcal{M}<4$).

In addition, we can also consider contributions from fossil CRes.
Particle-in-cell simulations have predicted that the acceleration efficiency ($\xi_{\rm e}$) of weak shocks can drastically increase with the presence of fossil CRes \citep[e.g.,][]{2015ApJ...809..186K}. 
This scenario is supported by the presence of the discrete radio source `I' in NR. 
Thus, the observed surface brightness ratio might be due to a larger amount of fossil CRes in NR.
\begin{figure}
    \centering
	\includegraphics[width=\columnwidth]{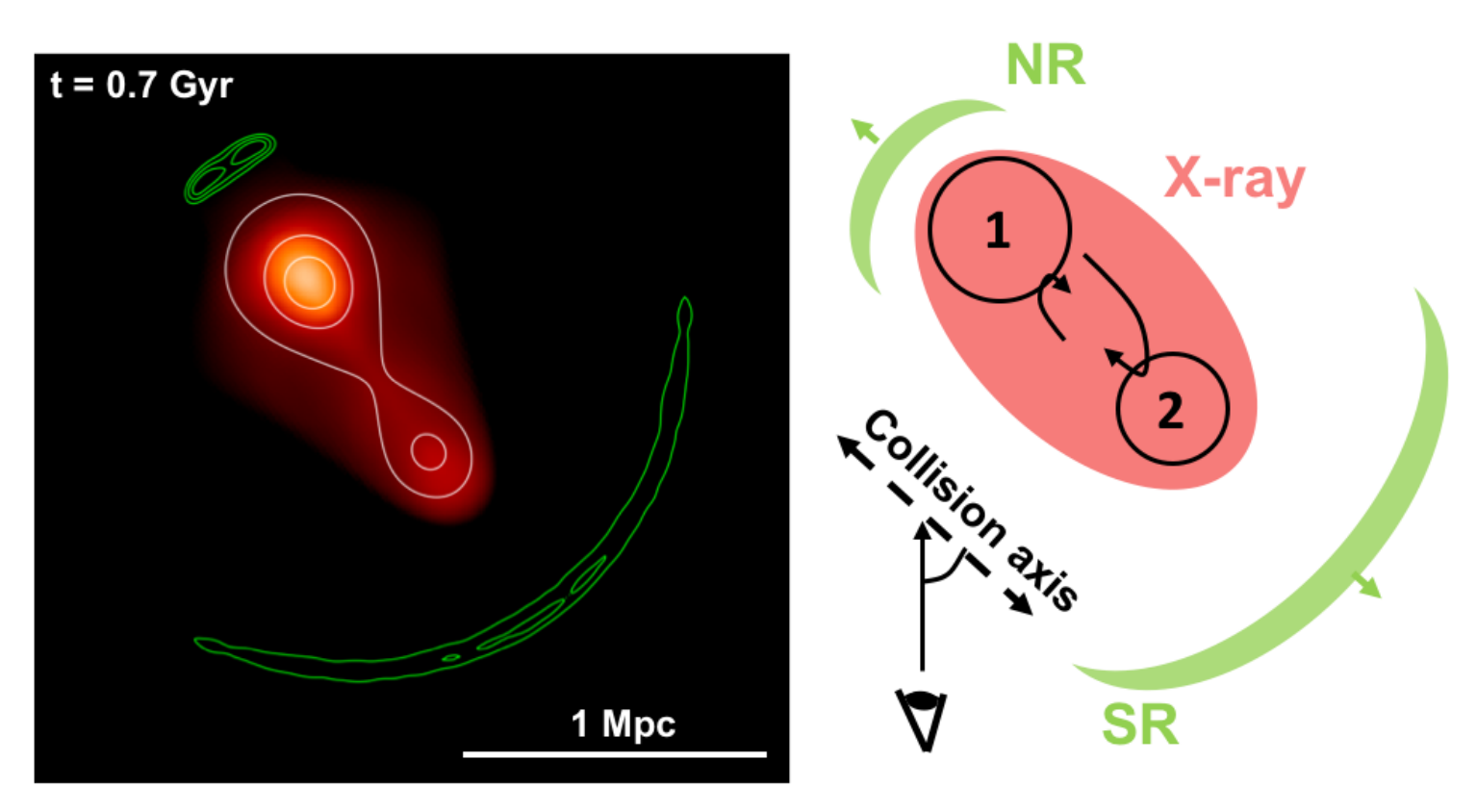}
    \caption{Simulated cluster merger at $\mytilde0.7\rm~Gyr$ since the first passage (Left) and its schematic description (Right). Projected mass contours (white) and kinetic flux contours (green) are overlaid over X-ray map (colormap) and the mass (kinetic flux) contours are smoothed with a $\mytilde50\rm~kpc$ ($\mytilde20\rm~kpc$) Gaussian kernel. The kinetic flux map is considered as the proxy of diffuse radio map. We can notice followings observed properties are reproduced with simulation: elongated X-ray map, lack of X-ray-mass dissociation, and asymmetric radio relics with a larger relic in front of the less-massive cluster.}
    \label{fig:scenario}
\end{figure}

\subsection{Investigating the Merger History of ZwCl1447 through Numerical Simulations}
\label{sec:merger_history}
Our multi-wavelength analysis of ZwCl1447 suggests that we are witnessing a NE-SW post-merger. Here, we further test our merger scenario using hydrodynamical simulations. Although one can compare a number of features between simulation and observation, our current focus is to produce the current positions of the radio relic, X-ray peaks, and mass concentrations.


We design and analyze simulations following the procedure in \cite{Lee2020}. To briefly summarize, we placed two spherically symmetric clusters along the `x'-axis, each consisting of an NFW dark halo and a $\beta-$profile ICM. 
The mass of each subcluster is provided by our WL analysis (\textsection\ref{sec:wl_mass}). We normalized the gas density in such a way that the baryon fraction becomes $\mytilde13\%$ at $R_{200}$. 
We added a small non-zero velocity in the `y'-axis direction to simulate a slight off-axis collision.
We ran the simulation using the adaptive mesh refinement (AMR) code {\tt RAMSES} \citep{Teyssier2002} and resolved the cells from the coarsest level of $\mytilde150\rm~kpc$ to the finest level of $\mytilde10\rm~kpc$ based on their density. The results were projected along the `y'-axis
onto the `xz'-plane. The initial non-zero velocity in the `y'-axis makes the viewing angle of the merger axis about 45\degr with respect to the plane of the sky.
We derived the projected maps of mass and X-ray as in \cite{Lee2020} and used the projected kinetic flux of the shock as a proxy for the radio map (see \citealt{Lee2020} for more details). 

Figure~\ref{fig:scenario} presents the simulation result at $\mytilde0.7\rm~Gyr$ after the first passage. 
At this epoch, the two subclusters turned around and are in the returning phase. The separation between the substructures in X-ray and mass is $\mytilde0.5\rm~Mpc$ similar to the observation. And the two merger shocks have travelled to large distances and their separation has reached $\mytilde1.8\rm~Mpc$, which is also similar to the current observed value. In addition, a larger shock is produced in front of the less massive southern cluster.
Therefore, this simulation shows that the observed features of ZwCl1447 can be explained with a binary cluster merger at $\rm TSC\sim0.7\rm~Gyr$.

The line-of-sight velocity is expected to be approximately $500\rm\,km\,s^{-1}$ at this epoch. However, because of the observational difficulty in identifying the cluster members associated with the southern cluster (\textsection\ref{sec:xray}), this comparison needs to await a deeper optical spectroscopic observation.
We expect that, with finetuning, more detailed properties (e.g., radio relics brightness) can also be reproduced. This is, however, beyond the scope of the current study.

The above TSC of roughly $0.7\rm~Gyr$ value from the simulation can be compared with the observed shock properties of the radio relics.
A shock propagation velocity is given by $\mathcal{M}c_{\rm s,1}$, where $c_{\rm s,1}$ is the sound speed in the upstream region.
In ZwCl1447, the average shock velocity is estimated as $\mytilde2,300\,\rm km\,s^{-1}$ when we assume that the upstream temperature is similar to that of the X-ray core ($T\sim3\rm~keV$).
If the two shocks are formed by the relative motion between the two subclusters and their propagation velocities are nearly constant, the current observed separation of $1.8\rm~Mpc$ yields $\rm TSC \sim0.8\rm~Gyr$, which is in good agreement with the simulation.

\section{Summary}
\label{sec:summary}
We have studied the unique merging galaxy cluster ZwCl1447 with multi-wavelength observations. Our new radio observations with uGMRT discovered a remarkable double radio relic in ZwCl1447, which
have not been resolved in previous studies.
Our main findings can be summarized as the followings:

\begin{itemize}
    \item  The southern radio relic is long ($\mytilde1.2\rm~Mpc$) and thin ($\mytilde0.1\rm~Mpc$), similar to the morphological features of the giant Sausage radio relic. The northern relic is short ($\mytilde0.3\rm~Mpc$).
    Our rdaio spectral analysis shows that weak shocks ($\mathcal{M}\lesssim3$) are associated with both relics.
    
    \item We discovered two additional diffuse radio structures (NCR and SCR), which are about $0.2\rm~Mpc$ closer to the cluster center than the adjacent relics. Since they do not appear to be associated with the cluster X-ray emission, the features are not likely to be part of radio halos. The presence of the discrete radio sources with optical cluster galaxy counterparts within the emission hints at the possibility that their origins might be related to star-formation activities.

    \item We found optical counterparts of most discrete radio sources and confirm that they are cluster galaxies. These radio sources featured $100\rm kpc$-size extended emissions along the merger axis, which hint at the fossil CRes injection to the ICM.
    
    \item The radio morphology and spectral profile suggest that the northern radio relic might be the result of re-acceleration of the fossil cosmic-ray electrons from the cluster galaxy, which is a discrete radio source at the eastern edge of the relic.
    
    \item Each relic follows the size-luminosity relation of other known relics in the literature. However, the the surface brightness ratio between the relics in ZwCl1447 is a significant outlier. 
    Possible causes might be the acceleration efficiency that decreases precipitously in the weak shock regime ($\mathcal{M}\sim3$) and/or the brightness boost in the northern relic due to re-acceleration.
    
    \item Our X-ray and weak lensing analysis found two consistent substructures that are aligned perpendicular to the double radio relic, supporting the NE-SW merger scenario inferred by our newly revealed radio relics.
    The total mass of ZwCl1447 is estimated to be about $5\times10^{14}M_{\rm sun}$ with a mass ratio of $3:1$ ($2.7\pm0.8\times10^{14}M_{\sun}$ and $1.0\pm0.5\times10^{14}M_{\sun}$ for the northern and southern halos, respectively.)
    
    \item Our numerical simulation reproduces the observed locations of the X-ray, weak-lensing mass, and radio relics. This shows that ZwCl1447 is currently at the returning phase and it experienced a near head-on collision about $0.7\rm~Gyr$ ago. 
    
\end{itemize}


Overall, despite our short on-source time, our uGMRT observations have resolved the radio morphology in ZwCl1447 and further found a peculiar surface brightness difference between the double radio relic. Future deep radio observations at the same and/or at different frequencies will better quantify the cluster properties and help us to better understand the shock acceleration theory.

We thank Nathan Golovich for sharing his Keck/DEIMOS optical spectrum data and
Juheon Lee and Ishwara Chandra for useful discussions.
MJJ acknowledges support for the current research from the National Research Foundation (NRF)
of Korea under the programs 2017R1A2B2004644 and
2020R1A4A2002885. RK acknowledges the support of the Department of Atomic Energy, Government of India under project no. 12-R\&D-TFR-5.02-0700. We thank the staff of the GMRT who have made these observations possible. The GMRT is run by the National Centre for Radio Astrophysics of the Tata Institute of Fundamental Research. The scientific results reported in this article are based on observations made by the Chandra X-ray Observatory 



\bibliography{reference}

\end{document}